\documentclass[aps,preprint]{revtex4}
\usepackage{graphicx}

\begin{document}

%\tightenlines

\title{$D$-mesons: In-medium effects at FAIR}

\author{L. Tol\'os$^{1,2,3}$\footnote{ e-mail:l.tolos@gsi.de}, J. Schaffner-Bielich$^2$ and  H. St\"ocker$^{2,3}$}

\affiliation{$^1$ Gesellschaft f\"ur Schwerionenforschung,
Planckstrasse 1, 64291 Darmstadt, Tel:+49-6159-71-2751, Fax:+49-6159-71-2990 \\
$^2$ Institut f\"ur Theoretische Physik, 
Johann Wolfgang Goethe Universit\"at, Max-von-Laue 1, 60438 
Frankfurt am Main \\
$^3$ FIAS, 
Johann Wolfgang Goethe Universit\"at, Max-von-Laue 1, 60438
Frankfurt am Main}

\date{\today}

\begin{abstract}

The $D$-meson spectral density at finite temperature is obtained within a self-consistent coupled-channel approach. For the bare meson-baryon interaction, a separable potential is taken, whose parameters are fixed by the position and width of the $\Lambda_c (2593)$ resonance. The quasiparticle peak stays close to the free $D$-meson mass, indicating a small change in the effective mass for finite density and temperature. Furthermore, the spectral density develops a considerable width due to the coupled-channel structure. Our results indicate that the medium modifications for the $D$-mesons in nucleus-nucleus collisions at FAIR (GSI) will be dominantly on the width and not, as previously expected, on the mass.

\vspace{0.5cm}

%13.75.Jz, 25.75.-q,21.30.Fe, 21.65.+f, 12.38.Lg, 14.40.Ev, 25.80.Nv


\end{abstract}

\maketitle

\section{Introduction}
\label{sec:intro}

Understanding the properties of matter under extreme conditions of density and temperature is 
a topic of high interest over the last years due to the direct implications in heavy-ion 
experiments  as well as in the study of astrophysical compact objects like
neutron stars. The CBM experiment of the FAIR project (Facility for Antiproton and Ion Research) at GSI (Darmstadt) \cite{CBM} will investigate highly compressed nuclear matter, permitting the exploration of the QCD phase diagram in the region of high-baryon densities. In particular, the research program will be focused on obtaining the in-medium modifications of hadrons in dense matter, such as mesons containing charm or anti-charm quarks \cite{letterofintent,Senger}.

The medium modifications of mesons with charm, such as $D$ and $\bar D$ mesons, have been object of recent theoretical interest
\cite{qmc,arata,weise,friman,dmeson,Tolos04} due to the consequences for J/$\Psi$ suppression, as observed at SPS energies by the NA50 collaboration \cite{NA501}. Furthermore, changes in the properties of the $D$/$\bar{D}$ mesons will have a strong effect on the predicted open-charm enhancement in nucleus-nucleus collisions \cite{kostyuk,cassing}. The open-charm enhancement  was introduced in order to understand the enhancement of 'intermediate-mass dileptons' in Pb$+$Pb collisions at the SPS energies. In Ref.~\cite{cassing} a suppression of $D$-mesons by a factor of 10 relative to a global $m_T$-scaling with a slope of 143 MeV was expected at bombarding energies of 25 AGeV. The inclusion of  attractive mass shifts lead to an enhancement of open-charm mesons by about a factor of 7 such that an approximate $m_T$-scaling was regained. However, the open-charm enhancement is a matter of recent debate, since the latest results on dimuon production by the NA60 collaboration are not consistent with an enhancement of open charm \cite{NA60}.
 
A phenomenological estimate based on the quark-meson coupling (QMC) model predicts an attractive $D^+$-nucleus potential at normal nuclear matter density $\rho_0$ of the order of -140 MeV \cite{qmc}. The $D$-meson mass shift has also been studied using the QCD sum-rule (QSR) approach \cite{arata,weise}. 
Due to the presence of a light quark in the $D$-meson, the mass modification of
the $D$-meson has a large contribution from the light quark condensates. A mass shift of -50 MeV at $\rho_0$ for the $D$-meson has been suggested \cite{arata}. A second analysis, however, predicts only a splitting of $D^+$ and $D^-$ masses of 60 MeV at $\rho_0$ because the uncertainties to which the mass shift is subject at the level of the unknown $DN$ coupling to the sector of charmed baryons and pions \cite{weise}. The mass modification of the $D$-meson is also addressed using a chiral effective model in hot and dense matter \cite{dmeson}. Similar results to the previous works on QMC model or QSR are obtained at T$=$0 with  the interaction Lagrangian of chiral perturbation theory.  
However, a larger mass drop is observed ($\sim$-200 MeV) at T$=$0 when a SU(4) chiral effective model is used. In this case, the attraction is reduced to $\sim$-150 MeV at T$=$150 MeV.

In all these investigations, the $D$-meson spectral density in dense matter is not studied. In our previous work \cite{Tolos04}, the $D$-meson spectral density is obtained by including coupled-channel effects as well as the dresssing of the intermediate propagators, which are not considered in the QMC models, the QSR approach or the chiral effective models. These effects turn out to be crucial for describing the $D$-meson in dense matter \cite{Tolos04}, as already pointed out for the $\bar K$ meson in the nuclear medium \cite{Lutz,Oset,Tolos01,Tolos02}. Thus, the attraction felt by the $D$-meson is strongly reduced or becomes slight repulsion in  Ref.~\cite{Tolos04}. A most recent coupled-channel calculation, appeared after submission of the original manuscript, supports this notion by finding a repulsive potential for the $D$-meson in dense matter \cite{Lutz05}. 

In this letter, the spectral density for the $D$-meson not only at finite density but also at finite temperature is studied using a self-consistent coupled-channel  $G$-matrix  calculation. The medium effects at finite temperature like the Pauli blocking on the nucleons or the Bose distribution on the pionic intermediate states, and the dressing of the $D$-mesons, nucleons and pions are investigated. Our calculation indicates that the in-medium properties of $D$-mesons in heavy-ion collisions will be dominantly on the width and not on the mass, contrary to previous expectations \cite{cassing}. Therefore, open-charm enhancement will only be observed at FAIR if the off-shell behaviour of the $D$-mesons in this hot and dense matter is taken into account.

\section{The $D$-meson spectral density at finite temperature}
\label{sec:formalism}

In this section, we extend our self-consistent calculation for the self-energy of a $D$-meson in infinite symmetric nuclear matter \cite{Tolos04} to finite temperature. We calculate 
the single-particle potential of the $D$-meson in symmetric nuclear matter at finite temperature 
by performing a self-consistent coupled-channel calculation taking, 
as the  bare $D N$ interaction, a separable potential model for s-wave.
The parameters of this model, i.e, coupling constant and cutoff, are 
determined by fixing the position and the width of the $\Lambda_c(2593)$ resonance. 
In our calculation we have assumed that our interaction is SU(3)
symmetric for u-,d- and c- quarks. This approach is similar to use SU(4)
symmetry and ignore the channels with strangeness which are higher in
mass and well above threshold. Using symmetry for the interactions,
even for very heavy particles, has been frequently used for charmed
hadrons before. It is motivated by the fact that the decay constants for
pions, kaons and $D$-mesons differ by only a factor two, while the mass of
the $D$-meson is about an order of magnitude larger than the pion mass. 
For quark masses the corresponding increase of mass is even two orders
of magnitude. Also, the mass splitting for the pseudoscalar and scalar
light, strange and charm quark sector is very similar. 

In a recent publication \cite{Hofman}, it was shown
that the ansatz of SU(4) symmetry for the meson interactions results in a generalized KSFR relation \cite{KSFR}.  Furthermore, a previous work on heavy-light meson resonances, i.e. meson resonances with open charm or bottom, was
performed in terms of the chiral quark model with a non-linear
realization of the chiral SU(3) symmetry of QCD \cite{Kolomeitsev}.
SU(4) symmetry for charmed hadrons was used by Ref.~\cite{Liu2002} to
describe J/Psi absorption on nucleons and found to be in agreement with
experimental data.  On the other hand, a SU(4) chiral linear sigma model was constructed in Ref.~\cite{Joerg} fitting to charmed mesons quite succesfully. In our approach, the parameters fitted to the $\Lambda_c(2593)$ are close to the ones for the fit to the $\Lambda(1405)$, supporting the notion of SU(4) symmetry for the interaction.
 
The introduction of temperature in the $G$-matrix calculation
affects the Pauli blocking of the nucleons or the Bose distribution on the pionic intermediate states together with the dressing of $D$-mesons, nucleons and pions. The $G$-matrix equation at finite $T$ reads formally as 
\begin{eqnarray}
&&\langle M_1 B_1 \mid G(\Omega,T) \mid M_2 B_2 \rangle = \langle M_1 B_1
\mid V \mid M_2 B_2 \rangle  +\nonumber \\
&&\sum_{M_3 B_3} \langle M_1 B_1 \mid V \mid
M_3 B_3 \rangle \frac {F_{M_3 B_3}(T)}{\Omega-E_{M_3}(T) -E_{B_3}(T)+i\eta} \langle M_3
B_3 \mid
G(\Omega,T)
\mid M_2 B_2 \rangle \ ,
   \label{eq:gmat1}
\end{eqnarray}
where $V$ is the separable potential and $\Omega$ is the  starting energy. In
Eq.~(\ref{eq:gmat1}),  $M_i$ and $B_i$  represent, respectively,
the possible mesons ($D$,$\pi$,$\eta$) and baryons ($N$,$\Lambda_c$,$\Sigma_c$), and their corresponding quantum numbers,
such as coupled spin and isospin, and linear momentum.  The
function $F_{M_3 B_3}(T)$ for the $D N$ states stands for the Pauli operator, i.e  $Q_{D N}(T)=1-n(k_N,T)$, where
$n(k_N,T)$ is the nucleon Fermi distribution at the corresponding
temperature. For $\pi \Lambda_c$ or $\pi \Sigma_c$ states, the function $F_{M_3 B_3}(T)$ is
$1+n(k_{\pi},T)$, with $n(k_{\pi},T)$ being the Bose distribution of pions at a given temperature. 
The function $F_{M_3 B_3}(T)$ is unity for the other intermediate states.

Temperature also affects the in-medium properties of the different particles involved. For nucleons, we have used a temperature-dependent Walecka-type $\sigma-\omega$ model with density-dependent scalar ($g_s$) and vector coupling ($g_v$) constants. These parameters are obtained from  Table 10.9 of Ref.~\cite{Machleidt}. For given $g_s$ and $g_v$ at $T=0$, and fixing the density $\rho$, the nucleonic spectrum
\begin{eqnarray}
E_N(k_N,T)=\sqrt{k_N^2+m^*(T)^2}-\Sigma^0 \ ,
\end{eqnarray}
with the vector potential $\Sigma_0$ and the effective mass $m^*$ defined as
\begin{eqnarray}
\Sigma^0&=&-\left(\frac{g_v}{m_v}\right)^2 \rho    \nonumber \\
m^*(T)&=&m-\left(\frac{g_s}{m_s}\right)^2\frac{4}{(2\pi)^3} \int
d^3k_N \ \frac{m^*(T)}{\sqrt{k_N^2+m^*(T)^2}} \  n(k_N,T) \ ,
\end{eqnarray}
and the chemical potential $\mu$, which is calculated imposing the normalization property at each density
\begin{eqnarray}
\rho=\frac{4}{(2\pi)^3} \int  d^3k_N  \ n(k_N,T) \ = \frac{4}{(2\pi)^3} \int  d^3k_N \ \frac{1}{1+{\rm
e}^{\left(\frac{E_N(k_N,T)-\mu}{T}\right)}} \ ,
\label{eq:density}
\end{eqnarray}
are obtained simultaneously. 

In the case of pions, the self-energy in nuclear matter at finite temperature has been obtained following a previous calculation of the pion self-energy in nuclear matter \cite{Oset90,Ramos94} incorporating finite temperature effects (see Appendix of Ref~\cite{Tolos02}).  The self-energy in nuclear matter \cite{Oset90,Ramos94} is obtained adding to a small repulsive and constant $s$-wave part \cite{Sek83-Mei89}, the $p$-wave contribution coming from the coupling to 1$p$-1$h$, 1$\Delta$-1$h$ and 2$p$-2$h$ excitations together with short-range correlations. These correlations are mimicked by the Landau-Migdal parameter $g'$ taken from the particle-hole interaction described in Ref.~\cite{Oset82}, which includes $\pi$ and $\rho$ exchange modulated by the effect of nuclear short-range correlations. Furthermore, we have also incorporated the self-energy of a pion in the presence of a hot pionic gas following Ref.~\cite{Rapp}. However, for temperatures of the order of 120 MeV for which $D$-mesons will be produced at FAIR, this contribution has proven to be negligible compared to the self-energy in the nuclear medium and, therefore, will be omitted in our calculation. 

The $D$-meson  potential at a given temperature  is then
calculated according to
\begin{eqnarray}
 U_{D}(k_{D},E_{D},T)= \int d^3k_N \ n(k_N,T) \ \langle D N \mid
 G_{D N\rightarrow D N} (\Omega = E_N+E_{D},T) \mid D N \rangle \ .
\label{eq:self}
\end{eqnarray}
As for the case of T$=$0, this is a self-consistent problem for the $D$-meson potential, since the effective interaction $G$ depends on the $D$-meson single-particle energy, which in turn depends on the $D$-meson potential.

After achieving self-consistency for the on-shell value
$U_{D}(k_{D},E_D,T)$, the complete energy dependence of the
self-energy $\Pi_D(k_D,\omega,T)$ can be obtained by
\begin{equation}
\Pi_D(k_D,\omega,T)=2\sqrt{m_D^2+k_D^2} \, U_{D}(k_D,\omega,T) \ .
\label{eq:relation}
\end{equation}
This self-energy can then be used to
determine the
 $D$-meson single-particle propagator in the medium
\begin{equation}
D_{D}(k_{D},\omega,T) = \frac {1}{\omega^2-m_{D}^2-k_{D}^2
-2 \sqrt{m_{D}^2+k_D^2} U_{D}(k_{D},\omega,T)} \ ,
\label{eq:prop}
\end{equation}
and the corresponding spectral density is
\begin{equation}
S_{D}(k_{D},\omega,T) = - \frac {1}{\pi} \mathrm {Im\,} D_{D}(k_{D},\omega,T)\ .
\label{eq:spec}
\end{equation}
This simplified self-consistent scheme proved to be sufficiently good for the $\bar K$ case, as already shown in Refs.~\cite{Tolos01,Tolos02}.

\section{$D$-mesons at finite temperature: results and discussion}
\label{sec:results}

\begin{figure}[htb]
\vspace{1cm}
\centerline{
     \includegraphics[scale=0.6]{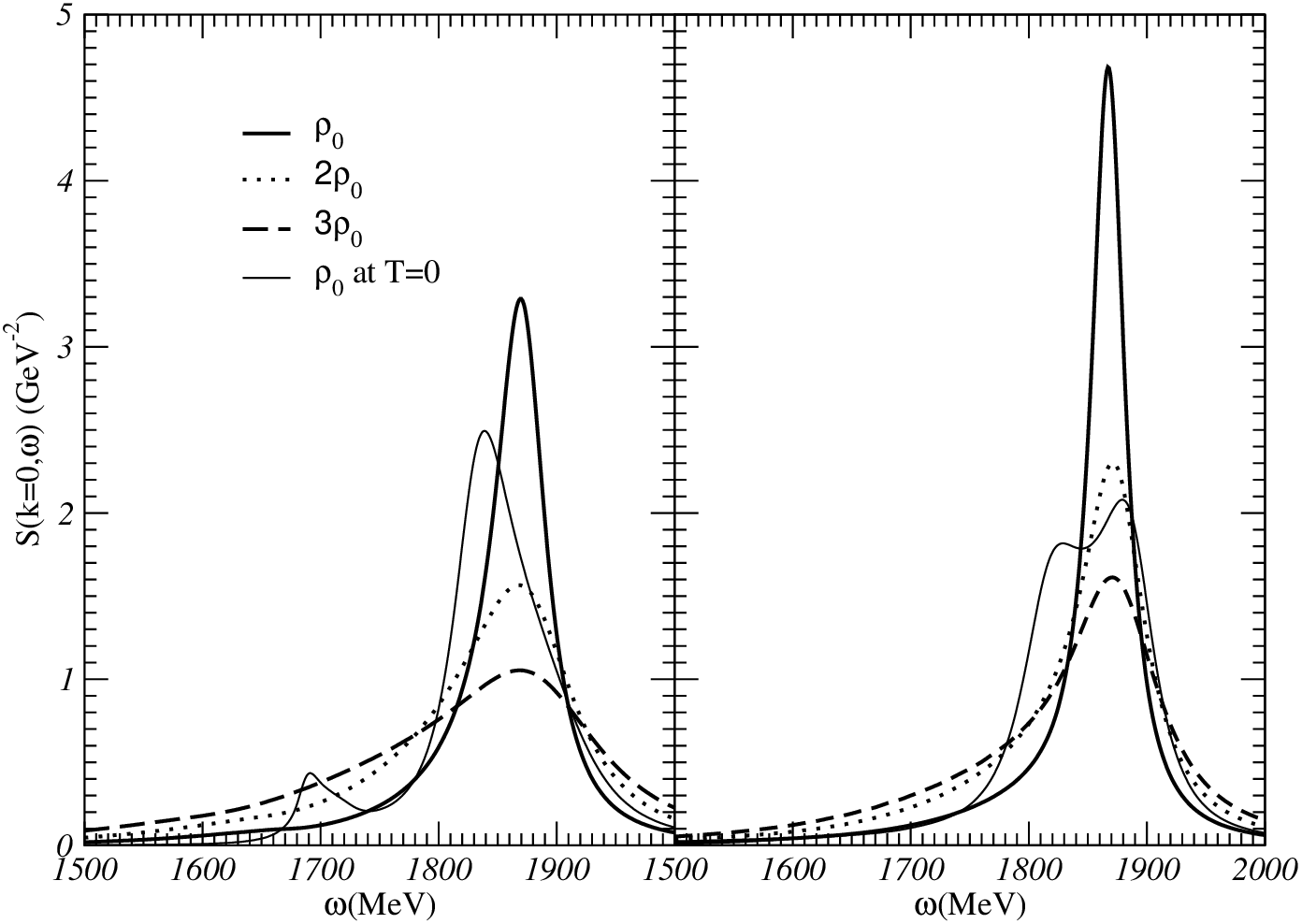}
}
      \caption{\small $D$-meson spectral density at $k_D=0$ and T$=$120 MeV as a function of energy for different densities, together with the $D$-meson spectral density at  $k_D=0$ and T$=$0 MeV for normal nuclear matter density $\rho_0$ in the two approaches considered: self-consistent calculation of the $D$-meson self-energy including the dressing of the nucleons in the intermediate states (left panel) and including not only the dressing of nucleons but also the self-energy of pions (right panel).}
        \label{fig:dmeson1}
\end{figure}

We start this section by discussing the effect of finite temperature in the $D$-meson spectral density. In Fig.~\ref{fig:dmeson1} the $D$-meson spectral density at zero momentum and T$=$120 MeV is shown as a function of the $D$-meson energy  for different densities and for a cutoff  $\Lambda=1$ and a coupling constant $g^2=13.4$. This is one of the sets of parameters that reproduce the position and width of the $\Lambda_c(2593)$ resonance (see Ref.~\cite{Tolos04}). The temperature is chosen in accord with the expected temperatures for which $D$-mesons will be produced at FAIR. The spectral density is displayed for the two approaches considered: self-consistent calculation of the $D$-meson self-energy including the dressing of the nucleons in the intermediate states (left panel) and the self-consistent calculation including not only the dressing of nucleons but also the self-energy of pions (right panel). The spectral density at zero temperature for nuclear matter saturation density, $\rho_0=0.17$ fm$^{-3}$, has also been included. 
In the case when only $D$-mesons and nucleons are dressed in the self-consistent process,
we observe an attraction of -23 MeV for $\rho_0$ at T$=$0, due to the attractive potential felt by the nucleon, which moves the $DN$ threshold to lower energies, i.e. closer to the $\Lambda_c(2593)$ resonance. For the full self-consistent calculation, the quasiparticle peak at T$=$0 stays close to the free position and mixes with a structure in the I$=$0 component of the in-medium $DN$ interaction, which lies close to the $DN$ threshold. This structure is a state with the $\Lambda_c$-like quantum numbers and can decay into $\pi \Sigma_c$ states, as already reported in Ref.~\cite{Tolos04}. In Fig.~4 of Ref.~\cite{Tolos04} this resonant structure was observed together with another $\Lambda_c$-like structure below the $\pi \Sigma_c$ threshold with decay channels $\pi(ph)\Sigma_c$ or $D(\Lambda_c h \pi)N$, where in parentheses  we have denoted an example for the component of the pion and $D$-meson that show up at energies below $\pi \Sigma_c$. This second smooth structure, however, is not visible in the spectral density because of lying far from the quasiparticle peak. The nature of these two resonances is definetely something that deserves further studies, being interesting to analize whether these two poles correspond to one effective resonance $\Lambda_c(2593)$ observed experimentally as in the case of the $\Lambda(1405)$ resonance \cite{Oset1405}.

Once finite temperature effects are included, the quasiparticle peak of the spectral density for $k_D=0$, defined as ${E_{qp}}^2=m_D^2+2 m_{D} {\rm Re} U_{D}(k_{D}=0,\omega,T)$, moves closer to the free mass. This expected behaviour is due to the fact that the Pauli blocking  is reduced with increasing temperature since the Fermi surface is smeared out with temperature. This behaviour was already reported for the case of $\bar K$ mesons \cite{Juergen,Tolos02}. Furthermore, structures present in the spectral distribution at T$=0$ are also washed out with increasing temperature. This is the case for the I$=$0 bump close to the quasiparticle peak  in the full self-consistent calculation (right panel) and the structure observed at energies around 1700 MeV for $\rho_0$, which corresponds to the $\Lambda_c(2593)$-hole excitation when only $D$-mesons and nucleons are dressed (left panel). This situation is similar to  the $\bar K N$ interaction, where the hyperon-hole excitations were also diluted at finite temperature \cite{Tolos02}. Note that only the $D$-meson spectral density at zero momentum has been shown, since a smooth dependence in momentum is obtained. Furthermore, the $D$-mesons produced in central nucleus-nucleus collisions have only small momenta in the rest frame of the hadronic fireball \cite{cassing}.

At finite density but zero temperature, works based on the QMC model \cite{qmc} and QCD sum-rules \cite{arata} predict an attractive $D^+$-nucleus potential with depths ranging from -50 to -140 MeV. These results are comparable to the ones obtained from an interaction lagrangian based on chiral perturbation theory, as reported in Ref.~\cite{dmeson}. However, in the same work,  when an effective chiral Lagrangian model is used, an even larger drop of the $D$-meson mass at T$=$0 is obtained ($\sim$-200 MeV). This investigation also addresses finite temperature effects, which reduce the attraction to -150 MeV for the latter case. In the present calculation, 
the coupled-channel effects, together with the self-consistent treatment of the $D$-meson properties, are responsible for the reduced attraction felt by the $D$-meson \cite{Tolos04}, or even repulsion as shown recently \cite{Lutz05}. In our model, the self-consistent coupled-channel effects result in an overall reduction of the in-medium modifications independently of the in-medium properties of the intermediate states \cite{Tolos04}. The previous works based on QCD sum-rules, QMC models or chiral effective lagrangians did not consider this coupled-channel structure and, therefore, do not generate dynamically the $\Lambda_c(2593)$ resonance. However, only a detailed comparison between models will be possible once the basic assumptions of each model, which are based on the unknown $DN$ interaction, will be constrained by the experimental data. Furthermore, finite temperature effects result in a less attractive $D$-meson  potential. At zero momentum, $D$-mesons at  $\rho_0$ and T$=$120 MeV interact with only a few MeVs of attraction for both approaches.

We have also obtained the decay width of $D$-mesons in nuclear matter, which has not been studied in previous mean-field calculations \cite{qmc,arata,weise,dmeson}. For T$=$120 MeV we observe a considerable broadening of the spectral density as density increases due to the  different decay channels. The strength of the spectral density at the quasiparticle peak is  reduced by a factor of about three when the density changes from  $\rho_0$ to $3 \rho_0$ in both approaches. Therefore, we conclude that the spectral density of $D$-mesons in a hot and dense medium develops a considerable width, while the quasiparticle peak stays close to the free mass. Note that 
 the width  has to be compared to the underlying
scale of the system, which would be the Fermi energy, the temperature
and the timescale of interactions in a heavy-ion collision, all of which
are of the order of 100 MeV.

\begin{figure}[htb]
\vspace{1cm}
\centerline{
     \includegraphics[scale=0.7]{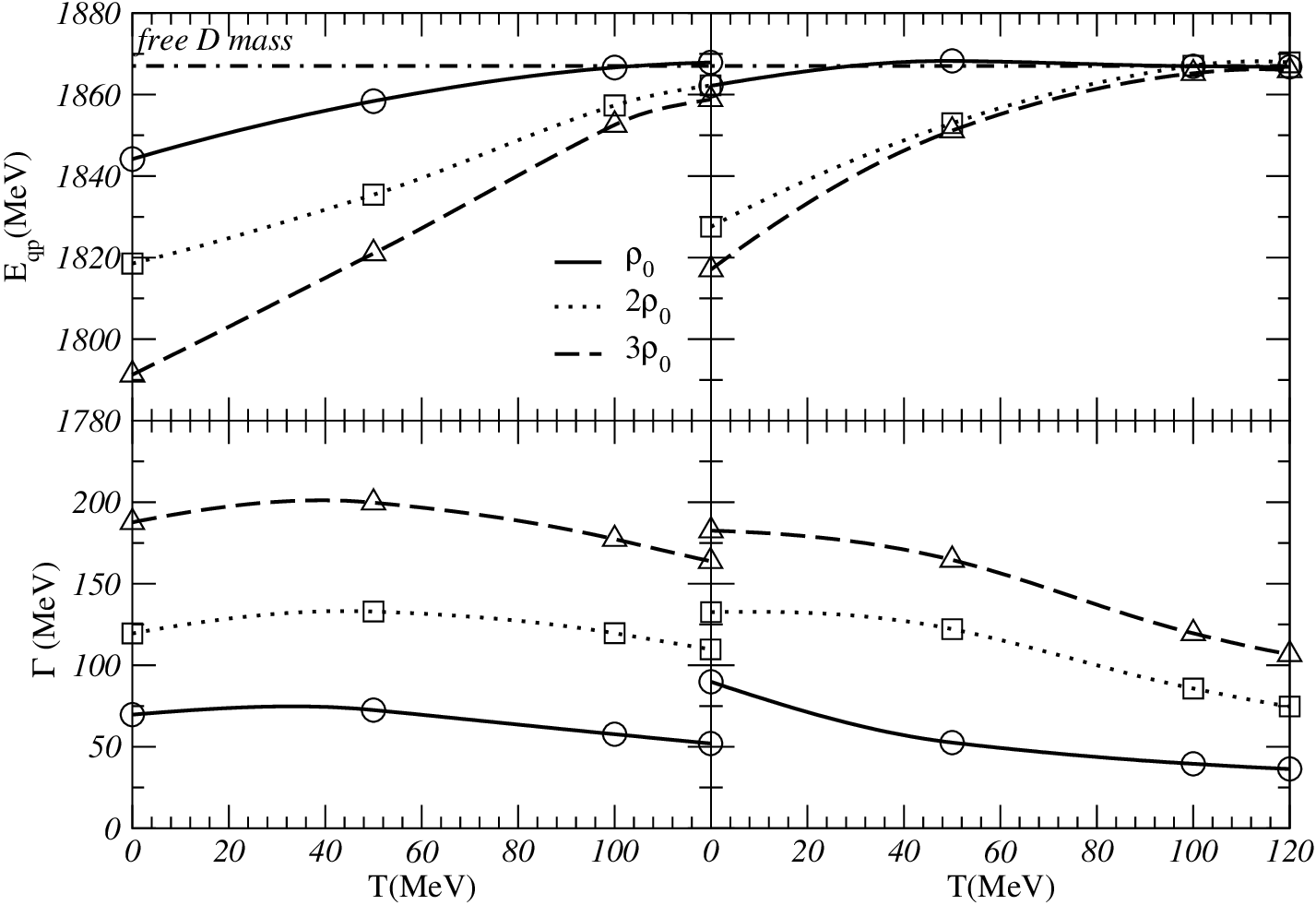}
}
      \caption{\small Quasiparticle energy and width of the $D$-meson spectral density at $k_D=0$ as a function of temperature for different densities and the two approaches considered: self-consistent calculation of the $D$-meson self-energy including the dressing of the nucleons in the intermediate states (left panels) and including not only the dressing of nucleons but also the self-energy of pions (right panels). 
}
        \label{fig:dmeson2}
\end{figure}

In order to better visualize the previous discussion, we plot in Fig.~\ref{fig:dmeson2} the evolution of  the quasiparticle peak together with the width of the $D$-meson spectral density at zero momentum as a function of the temperature for different densities (up to three times normal nuclear matter density) and for the  approaches  considered before. For zero temperature we observe a change of the $D$-meson mass with respect to its free value between -23 MeV for $\rho_0$  and -76 MeV for $3\rho_0$ when $D$-mesons and nucleons are dressed in the intermediate states (left panels). For the full self-consistent calculation (right panels), the $D$-meson potential lies between -5 MeV for $\rho_0$ and -48 MeV for $ 3 \rho_0$. For higher temperatures, the quasiparticle peak gets close to the $D$-meson free mass (at T$=$120 MeV and $\rho_0$, the peak is at -1 MeV for the first approach and -0.1 MeV for the second one).  We observe that the inclusion of the pion dressing reduces the attraction felt by the D-meson. A similar effect for antikaons was obtained \cite{Tolos04,Oset}. In the case of antikaons, the real part becomes less attractive and the imaginary part loses structure  once we include the pion dressing \cite{Tolos04}.

With regards to the width of the $D$-meson spectral density, at  T$=$120 MeV it increases from 52 MeV to 163 MeV for $\rho_0$ to $3\rho_0$ when $D$-mesons and nucleons are dressed. The increment goes from 36 MeV at $\rho_0$ to 107 MeV at $3 \rho_0$ for the full self-consistent calculation. Furthermore, we observe a  weak dependence on temperature when $D$-mesons and nucleons are dressed in the intermediate states. This behaviour can be understood by analyzing the evolution of the spectral density with temperature, which barely changes, on the left panel of Fig.~\ref{fig:dmeson1}. On the other hand, the width of the spectral density is slightly reduced with temperature when pions are dressed. At zero temperature, the quasiparticle peak mixes with a  I$=$0 $\Lambda_c$-like structure (right panel of Fig.~\ref{fig:dmeson1}). The inclusion of finite temperature washes out any structure and the resulting spectral function at finite temperature shows basically a single pronounced peak. This effect was already reported for the case of antikaons in Ref.~\cite{Tolos02}. Actually, as already observed for antikaons, we are recovering the $T\rho$ approximation with increasing temperature. In our calculation, pions do not couple directly to $D$-mesons and then, we do not take into account a collisional width with pions which is truely sensitive to temperature. 

We would also like to comment about the effect on  the $D$-meson self-energy due to the presence of a hot pion gas at the expected temperatures at FAIR for $D$-meson production (T $\approx$ 120 MeV). In Ref.~\cite{fuchs} the self-energy of a $D$-meson in a pion gas at finite temperature was calculated using the $D\pi$ forward resonance amplitude in isospin $1/2$, motivated by chiral symmetry \cite{Kolomeitsev}. For T$=$120 MeV, the potential felt by the $D$-meson in this hot pionic gas is of the order of -10 MeV while the width stays below 20 MeV. This effect should be added to our  $D$-meson self-energy in a hot baryonic medium shifting the quasiparticle peak by 10 MeV to lower energies and broadening the spectral density. However, a self-consistent calculation of the $D$-meson self-energy  in this pionic gas should be performed due to the presence of s-,p- and d-wave resonances for the $D$-$\pi$ system which can be easily modified in position and width in the presence of this hot pionic medium. 

The reduced attraction felt by the $D$-meson in hot and dense matter together with the large width observed  have important consequences for the $D$-meson production at the future CBM experiment at FAIR. Charm quark pairs are mainly created early in a heavy-ion collision by hard scattering. However, $D$-mesons can also be produced in secondary collisions after the primary hard NN collisions due to 'meson'-'baryon' rescattering via associated production (meson-baryon going to  $\Lambda_c$-$D$-meson, for example), as pointed out in Ref.~\cite{cassing}. Charm-exchange reactions to produce $D$-mesons are small, though. The associated production will be then modified in the hot and dense medium. In order to address the question to what extent in-medium modifications of the $D$-mesons could be experimentally seen, Ref.~\cite{cassing} studies the $m_T$-spectra for central $Au+Au$ collisions at 25 AGeV.  For this purpouse, a QCD sum-rule calculation of the $D$-meson mass shift \cite{arata} is used. The strong mass shift predicted by this QCD sum-rule calculation induces an enhancement of the $D$-meson yield by about a factor of 7 relative to the bare-mass case (see Fig.~16 of Ref.~\cite{cassing}).
 Our present calculation at T$=$0 predicts that the $D$-meson in nuclear matter feels a reduced attraction of -23 MeV at $\rho_0$ in the first approach and -5 MeV for the second one, which is half or even less, depending on the approach, of the $D$-meson mass shift obtained by  QCD sum-rules. Furthermore, finite temperature effects dramatically decrease this value for T$=$120 MeV. Therefore, if only the in-medium mass shift is considered, the $D$-meson yield is expected to lie very close to the free case. According to our model, the inclusion of a considerable width of the $D$-meson in the medium is the only source which could lead to an enhanced in-medium $D$-meson production, as studied for kaons in Ref.~\cite{Tolosratio}. 

\begin{figure}[htb]
\vspace{1cm}
\centerline{
     \includegraphics[scale=0.6]{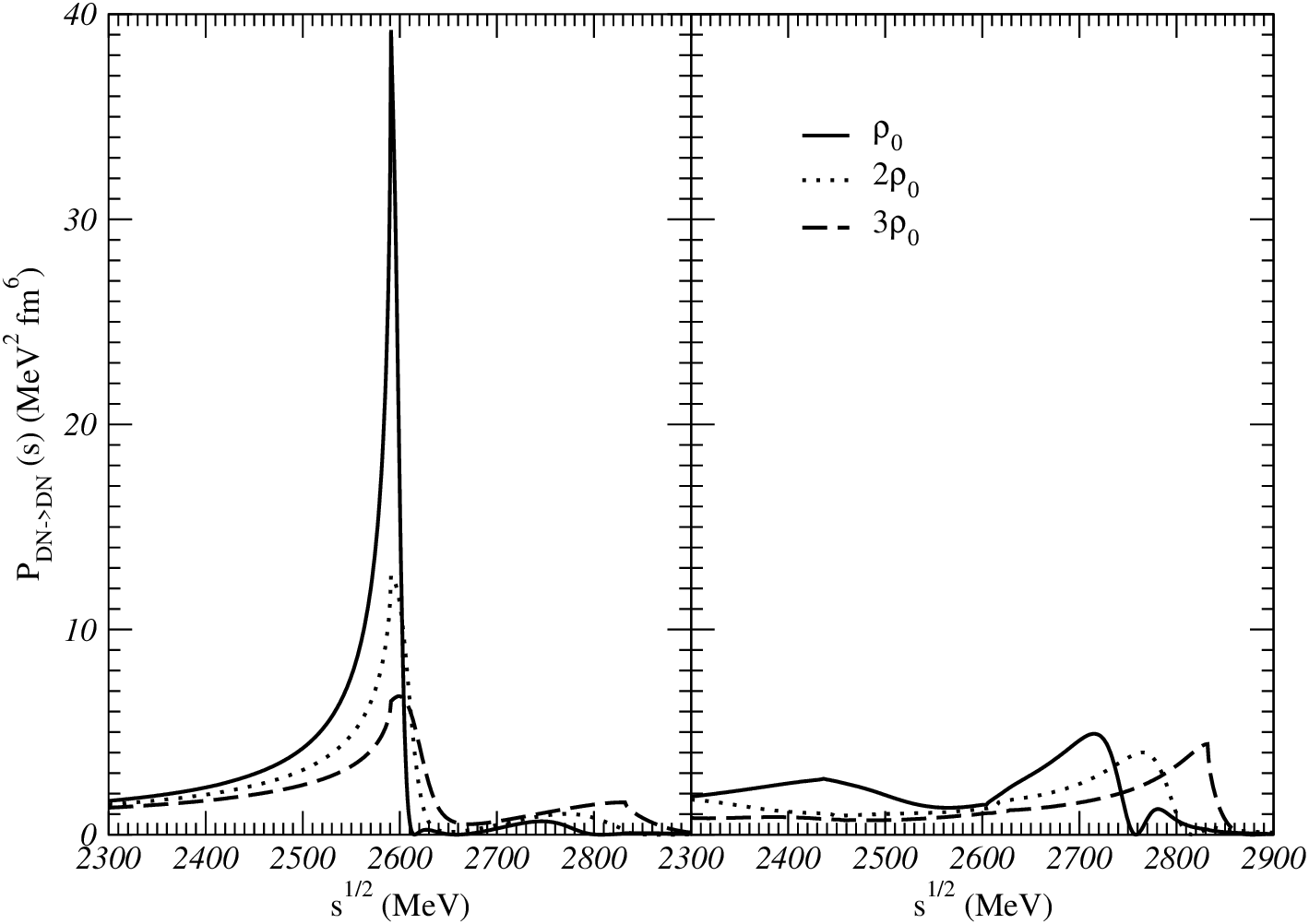}
}
      \caption{\small In-medium transition rates for $D^+n$ $(D^0 p)$ at T$=$120 MeV as a function of the center-of-mass energy for different densities and  the two approaches considered: self-consistent calculation of the $D$-meson self-energy including the dressing of the nucleons in the intermediate states (left panel) and including not only the dressing of nucleons but also the self-energy of pions (right panel). 
}
        \label{fig:dmeson3}
\end{figure}

The effect of the $\Lambda_c(2593)$ resonance for the $D$-meson subthreshold production as well as the magnitude of the elastic $D$-meson cross sections close to $DN$ threshold can be easily inferred from  Fig.~\ref{fig:dmeson3}. This figure shows the elastic in-medium transition rates for $D^+ n$ ($D^0 p$), the counterparts in the charm sector of $\bar K^0 n$ ($K^- p$), from $\rho_0$ to $3\rho_0$  at T$=$120 MeV as a function of the center-of-mass energy for the two approaches considered:  self-consistent calculation of the $D$-meson self-energy including the dressing of the nucleons in the intermediate states (left panel) and the self-consistent calculation including not only the dressing of nucleons but also the self-energy of pions (right panel). The in-medium transition rates are calculated from the off-shell in-medium $G$-matrix elements at finite temperature according to
\begin{eqnarray}
\label{crossp} P_{D_i+N_i \rightarrow D_f+N_f}(s) = 
\int d\cos(\theta) \
\frac{1}{(2s_{D_i}+1)(2s_{N_i}+1)} \sum_i \sum_\alpha \ G^\dagger G
\end{eqnarray}
where $D_{i,f}$ and $N_{i,f}$ represent the initial and final $D$-meson and nucleon states. The sums over $i$ and
$\alpha$ run over initial and final spins, while
$s_{D_i}, s_{N_i}$ are the spins in the entrance channel. 
Once the transition rates are known, the cross sections can be easily determined by multiplying with the phase space available according to
\begin{eqnarray}
 \sigma_{D_i+N_i \rightarrow D_f+N_f}(s) = 
\int d\cos(\theta) \
(2 \pi)^5 \frac{E_{N_i}\omega_{D_i} E_{N_f} \omega_{D_f}}{s} \frac{k'}{k} \frac{1}{(2s_{D_i}+1)(2s_{N_i}+1)} \sum_i \sum_\alpha \ G^\dagger G \
\end{eqnarray}
where $E$ and $\omega$ are the in-medium modified energies for the initial/final nucleon and $D$-meson, respectively, $k$ and $k'$ are the modulus of the momentum of the initial/final nucleon and $D$-meson in the center-of-mass frame and $s$ is the center-of-mass energy.

For the self-consistent calculation including the dressing of $D$-mesons and nucleons (left panel), we observe an enhanced transition rate for energies around the $\Lambda_c(2593)$ resonance mass. The maximum enhancement is reduced by a factor of about 4 as density grows from $\rho_0$ to $3 \rho_0$. This is a consequence of the dilution of the $\Lambda_c(2593)$ as the density is increased. However, when pions are also dressed in the self-consistent process (right panel), the enhanced transition rate is reduced drastically. As already discussed, the in-medium $DN$ interaction becomes smooth in the region where the $\Lambda_c(2593)$ resonance was lying and two resonant states with the $\Lambda_c$-like quantum numbers are generated for energies of 2.4-2.5 GeV and 2.7 GeV. The cross sections at threshold are expected on the order of 1 mb to 20 mb for the range of densities studied in both approaches.
Our findings are in contrast to what is obtained, for example, for the $K^-p$ elastic cross section which is of the order of 100 mb. In this case, the  $\Lambda(1405)$ resonance drives the behaviour of the cross sections at low momenta. 
Cross sections in the range of 10-20 mb for $D$,$D^*$ scattering with $\pi$ and $\rho$ were predicted by using an effective hadronic lagrangian which includes the charm mesons \cite{lin}. Based on that predictions, the HSD transport approach for nucleus-nucleus dynamics uses a constant cross section of 10 mb for $D$/$\bar{D}$ elastic scattering with mesons and also baryons. This is one of the ingredients in the collisional history of $D$,$\bar D$-mesons with baryons and mesons or quarks and diquarks in order to account for the open-charm production coming from secondary 'meson'-'baryon' reactions after the primary hard NN collisions. Further studies of the in-medium $D$-meson cross sections are required  in order to account for the $D$-meson production at FAIR.

\section{Conclusions and Outlook}
\label{sec:conclusions}

We have performed a microscopic self-consistent coupled-channel calculation of the spectral density of a $D$-meson embedded in symmetric nuclear matter at finite temperature, assuming a separable potential for the s-wave $DN$ interaction. The parameters for the separable potential, such as coupling constants and cutoffs, have been fitted to reproduce the position and width of the $\Lambda_c(2593)$ resonance.

The quasiparticle peak of the $D$-meson spectral density at finite temperature stays close to its free position for the range of densities analyzed (from $\rho_0$ up to $3\rho_0$) while the spectral density develops a considerable width. This small shift of the $D$-meson mass in the nuclear medium is in stark contrast with the large changes (-50 to -200 MeV) claimed in previous mean-field calculations, based on QCD sum-rules, QMC models or chiral effective lagrangians \cite{qmc,arata,weise,dmeson}.  In our model, the self-consistent coupled-channel effects result in an overall reduction of the in-medium modifications independently of the in-medium properties of the intermediate states. Furthermore, this self-consistent coupled-channel calculation also allows for the determination of the decay width of the $D$-meson, which we find around 40-50 $\rho/\rho_0$ for T$=$120 MeV.

The present coupled-channel approach to the $D$-meson properties in the nuclear medium at finite temperature is, as the first of its kind, exploratory and can be meliorated by improving the bare hadronic interaction. An extension of our SU(3) model for up, down and charm-quark content to a SU(4) model that incorporates other channels with strange-quark content  deserves further investigation. Studies in that direction have been recently performed saturating the interaction by a t-channel vector meson exchange \cite{Hofman,Lutz05}. Ref.~\cite{Lutz05}, which appeared after submission of the original version of this paper, substantiates our findings by obtaining even a repulsive mass shift and also a considerable decay width for $D$-mesons at zero temperature.

The in-medium effects devised in this work can be studied in heavy-ion experiments at the future International Facility at GSI. The CBM experiment will address, among others, the investigation of open charm \cite{letterofintent,Senger}. Our results imply that the effective masses of $D$-mesons, however, may not be drastically modified in dense matter at finite temperature, but $D$-mesons develop an important width in this hot and dense environment. Therefore, the abundance of $D$-mesons in nucleus-nucleus collisions should be calculated in off-shell transport theory. Our calculation indicates that the medium modifications to the $D$-mesons in nucleus-nucleus collisions will be dominantly on the width and not, as previously reported, on the mass.

\section*{Acknowledgments}

The authors are very grateful to A. Ramos and A. Kostyuk for critical reading of the manuscript, and P. Senger for useful discussions. L.T. wishes to acknowledge the financial support from the Alexander von Humboldt Foundation (AvH) and the Gesellschaft f\"ur Schwerionenforschung (GSI).

\newpage
%\setcounter{section}{2}
%\setcounter{equation}{0}

%\begin{references}

%\end{references}

%%%%%%%%%%%%% FIGURE 1 %%%%%%%%%%%%%%%%%%%%%%%%%%%%%%%%%%
%%%%%%%%%%%%%%%%%%%%%%%%%%%%%%%%%%%%%%%%%%%%%%%%%%%%%%%%%%%

\vspace{1cm} 

%%%%%%%%%%%%% FIGURE 2 %%%%%%%%%%%%%%%%%%%%%%%%%%%%%%%%%%
%%%%%%%%%%%%%%%%%%%%%%%%%%%%%%%%%%%%%%%%%%%%%%%%%%%%%%%%%%%

%%%%%%%%%%%%% FIGURE 3 %%%%%%%%%%%%%%%%%%%%%%%%%%%%%%%%%%
%%%%%%%%%%%%%%%%%%%%%%%%%%%%%%%%%%%%%%%%%%%%%%%%%%%%%%%%%%%


\begin{thebibliography}{999}

\bibitem{CBM} See http://www.gsi.de/fair/experiments/CBM

\bibitem{letterofintent} Letter of Intent for the Compressed Baryonic Matter Experiment at the Future Accelerator Facility in Darmstadt, Darmstadt, January 2004 (http://www.gsi.de/fair/experiments/CBM/5docu.html)

\bibitem{Senger} P. Senger, {\it Proceedings of the 30th International Workshop on Gross Properties of Nuclei and Nuclear Excitation: Hirschegg 2002: Ultrarelativistic Heavy Ion Collisions}, Hirschegg, Austria, 13-19 Jan 2002, published in *Hirschegg 2002, Ultrarelativistic heavy-ion collisions*, p. 55

\bibitem{qmc}
	K. Tsushima, D. H. Lu, A. W. Thomas, K. Saito, and R. H. Landau,
	Phys. Rev. C {\bf 59}, 2824 (1999);
	A. Sibirtsev,	K. Tsushima, and A. W. Thomas,
	Eur. Phys. J. A {\bf 6}, 351 (1999).
\bibitem{arata}
	A. Hayashigaki, Phys. Lett. B {\bf 487}, 96 (2000).
\bibitem{weise}
	P. Morath, W. Weise and S. H. Lee, {\it Proceedings of the 17th Autumm school on QCD: Perturbative or Nonperturbative?} Lisbon 1999, edited by L. S. Ferreira, P. Nogueira, and J. I. Silva-Marcos (World Scientific, Singapore, 2001), p. 425; W. Weise, {\it Proceedings of Hirschegg '01: Structure of Hadrons: 29th International Workshop on Gross Properties of Nuclei and Nuclear Excitations}, Hirschegg, Austria, 14-20 Jan 2001, published in *Hirschegg 2001, Structure of hadrons*, p. 249

\bibitem{dmeson}  A. Mishra, E. L. Bratkovskaya, J. Schaffner-Bielich,
S. Schramm and H. St\"ocker, Phys. Rev. C {\bf 69}, 015202 (2004).

\bibitem{friman}
	B. Friman, S. H. Lee and T. Song,
	Phys. Lett. B {\bf 548}, 153 (2002).

\bibitem{Tolos04} L. Tol\'os, J. Schaffner-Bielich and A. Mishra, Phys. Rev. C {\bf 70}, 025203 (2004).

\bibitem{NA501}
	M. Gonin {\it et al.}, NA50 Collaboration,
	Nucl. Phys. A {\bf 610}, 404c (1996).

\bibitem{kostyuk} M. I. Gorenstein, A. P. Kostyuk, H. St\"ocker and W. Greiner, Phys. Lett. B {\bf 509}, 277 (2001); M. I. Gorenstein, A. P. Kostyuk, H. St\"ocker and W. Greiner, J. Phys. G {\bf 27}, L47 (2001);  A. P. Kostyuk, M. I. Gorenstein and W. Greiner, Phys. Lett. B {\bf 519}, 207 (2001).


\bibitem{cassing}
	W. Cassing, E. L. Bratkovskaya, and A. Sibirtsev,
	Nucl. Phys. A {\bf 691}, 753 (2001).

\bibitem{NA60} Talk given at the XVIII International Conference on Nucleus-Nucleus Collisions (Quark Matter 2005) by  E. Scomparin (INFN, Torino, Italy), on behalf of the NA60 Collaboration. See http://qm2005.kfki.hu

\bibitem{Lutz} M. Lutz, Phys. Lett. B {\bf 426}, 12 (1998). 

\bibitem{Oset} A. Ramos and E. Oset, Nucl. Phys. A {\bf  671}, 481 (2000).

\bibitem{Tolos01} L. Tol\'os, A. Ramos, A. Polls, and T.T.S. Kuo,
Nucl. Phys. A {\bf 690}, 547 (2001).

\bibitem{Tolos02} L. Tol\'os, A. Ramos, and A. Polls, Phys. Rev. C {\bf 65}, 
054907 (2002).

\bibitem{Lutz05} M. F. M. Lutz and C. L. Korpa, nucl-th/0510006.

\bibitem{Hofman}J. Hofmann and M. F. M. Lutz, hep-ph/0507071.

\bibitem{KSFR}K. Kawarabayashi and M. Suzuki, Phys. Rev. Lett {\bf 16}, 255 (1966); Riazuddin and Fayyazuddin, Phys. Rev. {\bf 147}, 1071 (1966); D. Djukanovic et al., Phys. Rev. Lett. {\bf 93}, 122002 (2004).

\bibitem{Kolomeitsev} E. E. Kolomeitsev and M. F. M. Lutz, Phys. Lett. B {\bf 582}, 49  (2004).

\bibitem{Liu2002} W. Liu, C. M. Ko and Z. W. Lin, Phys. Rev. C {\bf 65}, 015203 (2002).

\bibitem{Joerg} D. Roder, J. Ruppert, D. H. Rischke, Phys. Rev. D {\bf 68}, 016003 (2003). 

\bibitem{Machleidt} R. Machleidt, Adv. Nucl. Phys. {\bf 19}, 189 (1989).

\bibitem{Oset90}E. Oset, P. Fernandez de Cordoba, L. L. Salcedo, and R. Brockmann, Phys. Rep. {\bf 188}, 79 (1990).

\bibitem{Ramos94}A. Ramos, E. Oset, and L. L. Salcedo, Phys. Rev. C {\bf 50}, 2314 (1994).

\bibitem{Sek83-Mei89}R. Seki and K. Masutani, Phys. Rev. C {\bf 27}, 2799 (1983);
O. Meirav, E. Friedman, R. R. Johnson, R. Olszewski, and P. Weber, Phys. Rev. C {\bf 40}, 843 (1989).

\bibitem{Oset82}E. Oset, H. Toki, and W. Weise, Phys. Rep. 83, {\bf 28} (1982).

\bibitem{Rapp} R. Rapp and  J. Wambach, Phys. Lett B {\bf 315}, 220 (1993).

\bibitem{Oset1405} D. Jido, J. A. Oller, E. Oset, A. Ramos and U. Meissner,  Nucl. Phys. A {\bf 725}, 181 (2003).

\bibitem{Juergen} J. Schaffner-Bielich, V. Koch and M. Effenberg, Nucl. Phys. A {\bf  669}, 153 (2000).

\bibitem{fuchs}C. Fuchs, B. V. Martemyanov, A. Faessler and M. I. Krivoruchenko, nucl-th/0410065.

\bibitem{Tolosratio} L. Tol\'os, A. Polls, A. Ramos and J. Schaffner-Bielich, Phys. Rev. C {\bf 68}, 024903 (2003).

\bibitem{lin} Z. Lin, and C. M. Ko, J.Phys. G {\bf 27}, 617 (2001).

\end{thebibliography}
\end{document}